\documentclass[sigplan,9pt,authorversion]{acmart}
\usepackage{subcaption}

\setcopyright{none}
\renewcommand\footnotetextcopyrightpermission[1]{} % removes footnote with conference information in first column

\begin{document}

\title{C for a tiny system}
\subtitle{Implementing C for a tiny system and making the architecture more suitable for C}

%\title{C for a tiny system\\ \small Implementing C for a tiny system and making the architecture more suitable for C}

\author{Philipp Klaus Krause, Nicolas Lesser}
%\author{Philipp Klaus Krause}
%\affiliation
%{
%	\institution{Albert-Ludwigs-Universität}
%	\city{Freiburg}
%	\state{Germany}
%}
%\author{Nicolas Lesser}
%\affiliation
%{
%	\institution{}
%	\city{}
%	\state{}
%}

\begin{abstract}
We have implemented support for Padauk microcontrollers, tiny 8-Bit devices with 60 B to 256 B of RAM, in the Small Device C Compiler (SDCC), showing that the use of (mostly) standard C to program such minimal devices is feasible. We report on our experience and on the difficulties in supporting the hardware multithreading present on some of these devices. To make the devices a better target for C, we propose various enhancements of the architecture, and empirically evaluated their impact on code size.
\end{abstract}




\maketitle

\pagestyle{plain} % removes running headers

\section{The architecture}

Padauk microcontrollers use a Harvard architecture with an OTP or Flash program memory and an 8 bit wide RAM data memory. There is also a third address space for input / output registers. These three memories are accessed using separate instructions.
Figure~\ref{architecture} shows the 4 architecture variants, which are commonly called pdk13, pdk14, pdk15 and pdk16 (these names are different from the internal names found in files from the manufacturer-provided IDE) by the width of their program memory. Each instruction is exactly one word in program memory. Most instructions execute in a single cycle, the few exceptions take 2 cycles. Most instructions use either implicit addressing or direct addressing; the latter usually use the accumulator and one memory operand and write their result into the accumulator or memory.
On the pdk13, pdk14 and pdk15, the bit set, reset and test instructions, which use direct addressing, can only access the lower half of the data address space.

\begin{figure}
\centerline{
\setlength\tabcolsep{2.5pt} % default value: 6pt
\hspace{-3mm}
\begin{tabular}{l||c|c|c|c}
subarchitecture & pdk13 & pdk14 & pdk15 & pdk16 \\
\hline\hline
internal name & SYM\_84B & SYM\_85A & SYM\_86B & SYM\_83A/B2 \\
prog.\ mem.\ width & 13 & 14 & 15 & 16 \\
prog.\ addr.\ bits & 10 & 11 & 12 & 13 \\
data addr.\ bits & 6 & 7 & 8 & 9 \\
I/O addr.\ bits & 5 & 6 & 7 & 6 \\
hardware threads & 1 & 1 or 2 & 1 & 2, 4 or 8 \\
\end{tabular}
}
\vspace{-0.5mm}
\caption{\label{architecture}Architecture variants}
\vspace{-3.5mm}
\end{figure}

There also are minor variations within the pdk14, pdk15 and pdk16 architectures in the form of some optional instructions only supported by some devices, e.g. the multiplication \texttt{mul} and the 16-bit wide push from data memory onto the stack \texttt{pushw m}.

The hardware multithreading support is implemented as a barrel processor similar to the Honeywell 800~\cite{Honeywell800} and the XCore~\cite{XCore}.
For each hardware thread, there is an 8 bit accumulator, an 8 bit stack pointer, a 4 bit flag register and a program counter. The manufacturer calls the combination of these 4 per-hardware-thread registers ``processing unit'', ``core'', ``FPP'' or ``FPPA''. While that terminology does not seem correct, as just one common ALU is shared among all of them, and during each clock cycle only one of them is active, we will still use the term ``core'' in the remainder of this document.

There is only one interrupt priority level and all interrupt handlers are executed on the first core.

For access to data memory, the only instructions with indirect addressing are \texttt{idxm a, m} and \texttt{idxm m, a}, that use a 16-bit-aligned address at data memory location \texttt{m} as address to read a byte into the accumulator or write a byte from the accumulator. For indirect access to program memory, the pdk15 and pdk16 have the \texttt{ldtabl a, m} and \texttt{ldtabh a, m} instructions that use a 16-bit-aligned address at data memory location \texttt{m} to read the lower or upper bits of a word in program memory into the accumulator. The pdk13 and pdk14 instead have the undocumented \texttt{ldsptl a} and \texttt{ldspth a} instructions that use the 16-bit address immediately above the top of the stack (i.e. at the location that will be immediately overwritten when an interrupt occurs) to read the lower or upper bits of a word in program memory into the accumulator.
%While the pdk16 architecture allows for up to 512 B of data memory, current devices are all in the range of 60 B to 256 B. - no longer correct due to recently released PFC886.

\begin{figure}
\centerline{
\includegraphics[width=\linewidth]{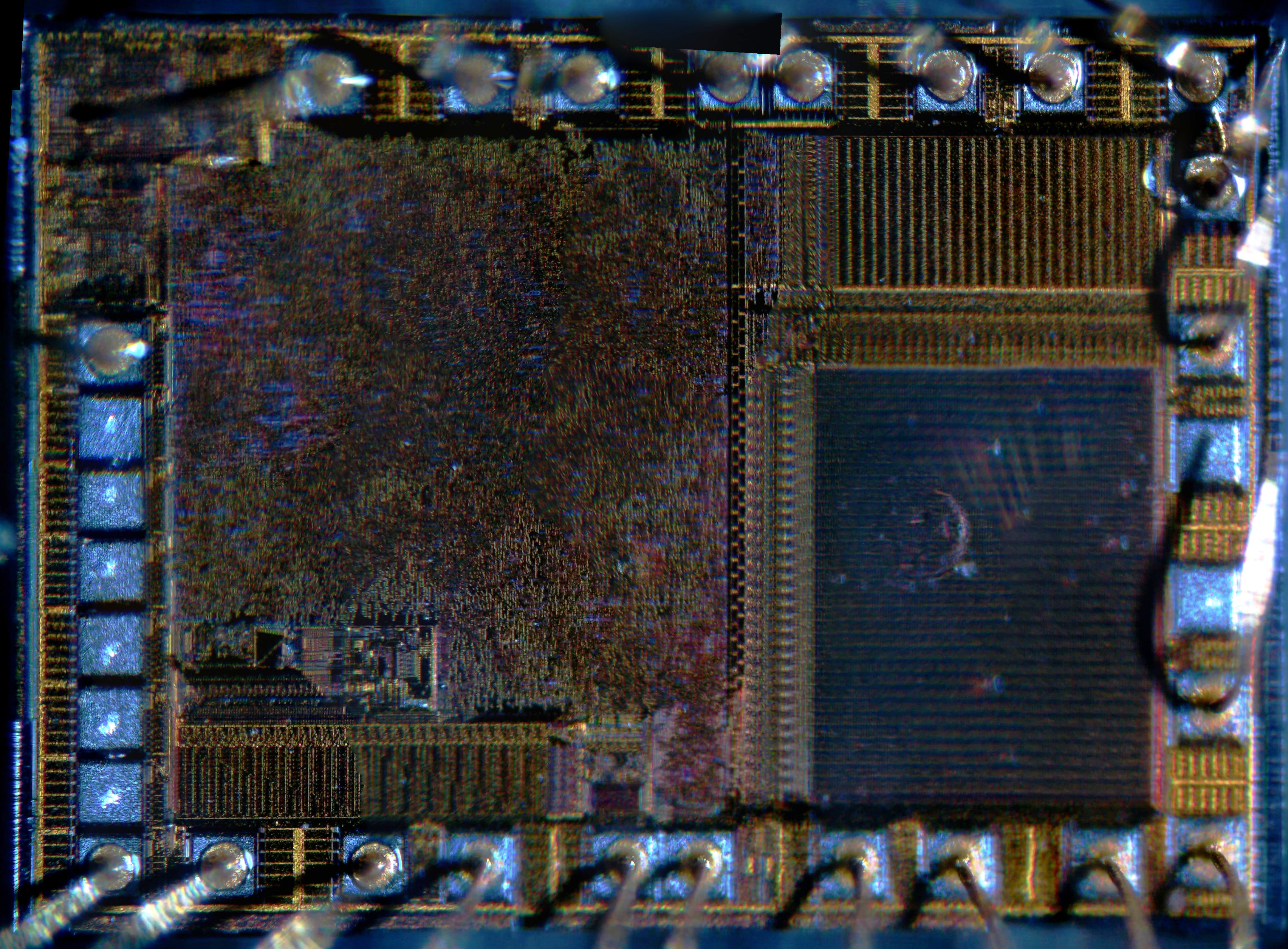}
}
\vspace{-0.5mm}
\caption{\label{PMC234-die}PMC234 die}
\vspace{-3.5mm}
\end{figure}

We decapped some devices using colophony~\cite{Beck1988}. Figure~\ref{PMC234-die} shows the die of the PMC234, a relatively large device with 4 kilowords of program memory and (compared to other Padauk µC) many peripherals. We can see that the area needed for the program memory (dark blue lower right area) including address decoders (neighbouring regular structure to the left and above) is about the same as for all the rest of the digital logic combined (irregular structures covering the center and areas towards the upper left from the center).

The architecture was clearly not designed to be program\-med in a high-level language. Padauk provides an IDE called ``Mini-C'', which, despite the name, uses essentially just assembler with some C-like syntax layered on top in a few places. We still wanted to find out, if and how such a minimal architecture could be a target architecture for a standard C compiler.

\section{Implementing C for a single core}

We implemented C for the 3 architecture variants, for which single-core devices exist: pdk13, pdk14 and pdk15. Our work consists of assemblers, linker support, simulators and backends in the Small Device C Compiler (SDCC)~\cite{SDCC}, has been merged upstream and is part of the SDCC 4.0.0 release. It is based on official documentation (which is somewhat incomplete, e.g. lacking opcodes and omitting a few instructions) as well as third-party information from reverse-engineering. Combined with the Easy PDK programmer hard-, firm- and software, our work forms a free~\cite{Stallman2002} toolchain targeting the Padauk µC. C standards are supported to the full extent that they are supported by SDCC (SDCC does not yet support standard compliant \texttt{double} and \texttt{long double} data types), there are no additional restrictions specific specific to the Padauk backend (but see the notes on reentrancy below).

Since there are no general-purpose registers, we use a single byte at memory address zero as a pseudo-register. While further pseudo-registers could help with code generation, they would also increase interrupt latency: all pseudoregisters have to be saved in the interrupt handler. Since 8-bit reads and writes are always atomic, the implementation of the most basic atomic type, \texttt{volatile sig\_atomic\_t} (a type for an atomic register with read and write access only) for communication with the signal handler is easy.

Access to the I/O memory is done by declaring variables there using special keywords (i.e. as extensions to C, as common in compilers targeting architectures where I/O is not memory-mapped).
However, that still leaves two address spaces to handle (placing all objects in data memory is not an option due to its small size) within the generic address space. We decided to use 16 bit wide pointers; the top bit, when set, indicates that the pointer points to code memory, and to data memory otherwise. Reads via pointers are translated into calls to a support function. Writes via pointers are simpler: we can assume the pointer to always point to data memory (otherwise the written object would have been declared \texttt{const}, resulting in undefined behaviour). We implemented an intraprocedural static analysis, that for simple cases can prove that a pointer points to data or code memory, resulting in optimizations such as pointer reads that do not need to call into support routines and narrowing pointers to data memory to 8 bits.
Objects in code memory are implemented using \texttt{ret k} instructions; to read a byte we call the address as if it was a function pointer. File scope objects that are declared \texttt{const} are placed in code memory, other objects are placed in data memory.

Due to lack of hardware support, multiplicative operators (with few exceptions) are translated into support routine calls.

The lack of a stack-pointer-relative addressing mode makes access to local variables on the stack inefficient, requiring explicit computation of stack offsets and reading via \texttt{idxm} (e.g. a 16 bit addition with operands on the stack typically needs 34 instructions / 40 cycles vs. just 6 instructions / 6 cycles for operands at fixed addresses). We therefore by default place local variables at fixed locations in memory instead (i.e. as if they were \texttt{static}), an approach common when targeting architectures that do not allow efficient access to the stack, but in violation of the C standard, as it makes function non-reentrant (in particular, recursive function calls are not possible with non-reentrant functions). Using the keyword \texttt{\_\_reentrant}, this can be changed for individual functions, or using the compiler option \mbox{\texttt{--stack-auto}} for whole compilation units.

\section{Implementing C for multiple cores}

Reentrancy is much more important on multicore devices, since functions could be executed simultaneously on multiple cores. Not only would functions have to be compiled as reentrant by default, we'd also need locks to protect the pseudoregister. This would come at a substantial runtime and code size cost.

Another way to avoid the reentrancy problem would be to have a copy for each function / core pair, and statically analyze the call graph to select the correct version at compile time. There would be one pseudoregister per core. This solution would be best for runtime, but would result in a substantial code size (and data size, due to duplication of local variables) increase, especially since it would also apply to the standard library and support functions.

Implementing atomics beyond \texttt{volatile sig\_atomic\_t} is complicated. While there is an atomic 8-bit swap instruction (which is useful to implement spinlocks), it only has a direct addressing mode, so it is not suitable even for \texttt{atomic\_flag} (a type with two possible values that can be read, written and swapped atomically). Since the C standard requires \texttt{atomic\_flag} to be ``lock-free'' (which, in the C standard means it can be used from signal handlers), there is no straightforward way to implement it using locks, assuming that we want interrupt handlers, which are always executed on the first core, to behave as signal handlers. We found a way to emulate ``lock-free'' atomics on the Padauk µC using two locks (when accessing the atomic the first core only takes the first lock, all other cores take the second lock first, then the first lock; the signal handler takes the second lock only, and can then check if the first lock is currently held; if it is held, we know that the signal handler accesses the atomic while the first core was in the middle of such an access and use special handling for that case). It is slow, which is problematic, since C programmers, from experience on other architectures, where ``lock-free'' atomics are really lock-free, tend to assume that ``lock-free'' atomics are very efficient.

Implementation of standard C mutexes and condition variables is unproblematic. Access to thread-local storage is slightly complicated by lack of an efficient way to find out on which core code is executing, requiring a binary search of the stack pointer value in the stack ranges assigned to cores. On the other hand, the C thread functions seem too heavyweight for such small devices; a lightweight approach of one startup function per core seems preferable.

\section{Improving the hardware}

We looked into small, incremental improvements to the architecture to make it a more suitable target for C. In the previous sections, inefficient stack handling was identified as the main weakness; we now consider two possible improvements: Additional instructions for stack handling and adding a stack-pointer-relative addressing mode.

To add instructions, we need existing gaps in the opcode map or instructions that can be dropped to create such gaps. Since all instructions have the same fixed width, the size of the gap needs to be big enough to include the operand. We found the pdk15 and pdk16 to have plenty of gaps in the opcode map, wide enough to add multiple instructions with memory operands. On the other hand, the pdk14 has no such gap; the two widest gaps are 88 and 67 subsequent opcodes (adding an instruction with a memory operand would require a gap of width at least 128). For the pdk13, there are only 3 gaps, of 35, 8 and 5 opcodes (adding an instruction with a memory operand would require a gap of width at least 64).
The \texttt{not a} instruction is redundant, as it is equivalent to \texttt{xor a, \#0xff}, but it is only a single opcode. The \texttt{addc m, a} instruction (it adds the current value in the accumulator and the carry bit to a byte in memory) is used rarely by the compiler and can be emulated (though not atomically) using \texttt{xch} and \texttt{addc a, m}. The \texttt{nadd m, a} and \texttt{nadd a, m} instructions are rarely used and easily emulated as well, but are present only on pdk15 and pdk16.

For easy adjustment of the stack pointer, we suggest using existing gaps to add an \texttt{spadd \#k} instruction, which would add its immediate operand to the stack pointer. Valid arguments \texttt{k} would be a subrange of the even numbers ($-8, \ldots, 6$ for pdk13, $-16, \ldots, 14$ for pdk14, more for pdk15 and pdk16).

For better access to the stack we looked into two options: The addition of \texttt{mov} variants that support stackpointer-relative addressing for a quarter of the address space (using existing gaps in the opcode map), and the use of the upper two address bits to switch from direct addressing mode to stackpointer-relative addressing. The latter option would allow all instructions that currently operate on memory to operate on the stack. But it would make a quarter of the address space inaccessible to direct addressing.

To evaluate the impact of these changes, we implemented support in a fork of SDCC 4.0.0 and used the smallest established integer benchmarks: Dhrystone 2.1~\cite{Dhrystone2}, Coremark 1.0~\cite{Coremark}, stdcbench 0.6~\cite{stdcbench}. Even these small benchmarks need more data memory than the Padauk µC have, and thus cannot be executed; we just use them to compare code size (since nearly all instructions are single-cycle, we expect runtime and energy use to closely follow code size). For this purpose we consider these established benchmarks to be more suitable than home-grown micro-benchmarks designed to fit into the data memory available.
The benchmarks were compiled for the pdk13, pdk14 and pdk15 architectures, and in two configurations: once only functions that actually need to be reentrant for correctness were manually marked and compiled as such, and once we compiled all code as reentrant.

\begin{figure*}
\centerline{
\subcaptionbox{\label{results:pdk13}pdk13}{\includegraphics{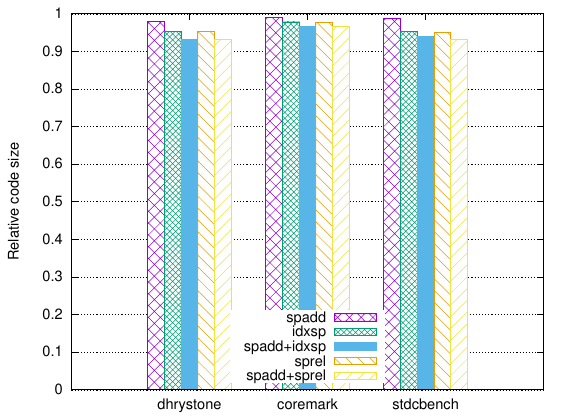}}
\subcaptionbox{\label{results:pdk13-r}pdk13, all reentrant}{\includegraphics{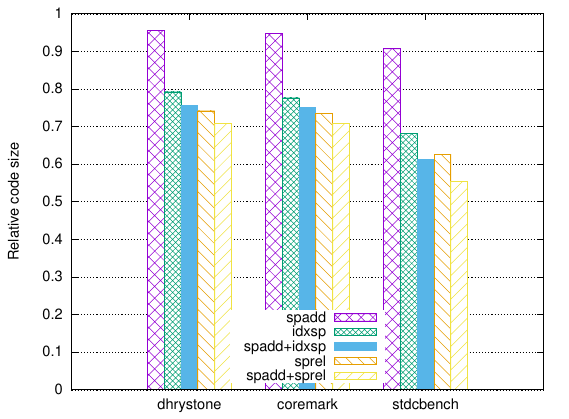}}
}
\centerline{
\subcaptionbox{\label{results:pdk14}pdk14}{\includegraphics{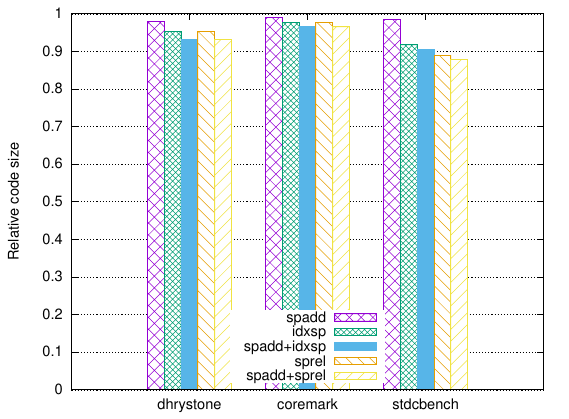}}
\subcaptionbox{\label{results:pdk14-r}pdk14, all reentrant}{\includegraphics{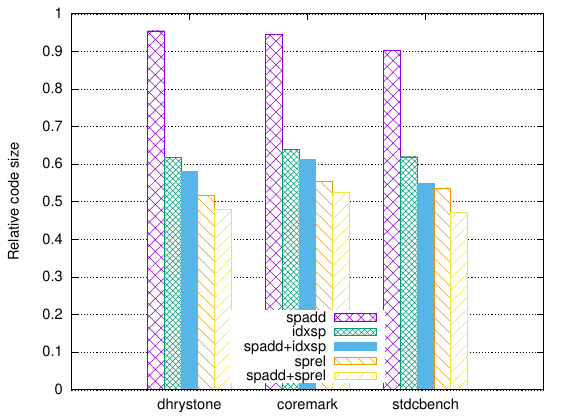}}
}
\centerline{
\subcaptionbox{\label{results:pdk15}pdk15}{\includegraphics{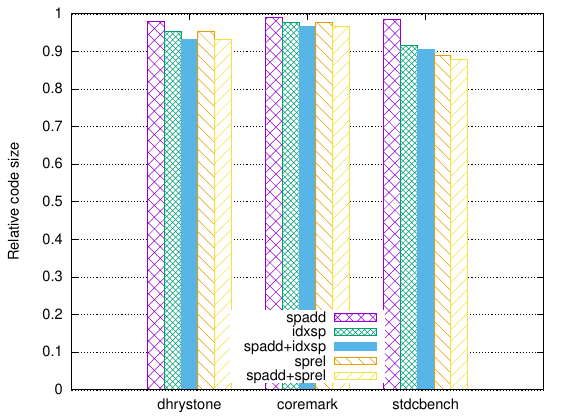}}
\subcaptionbox{\label{results:pdk15-r}pdk15, all reentrant}{\includegraphics{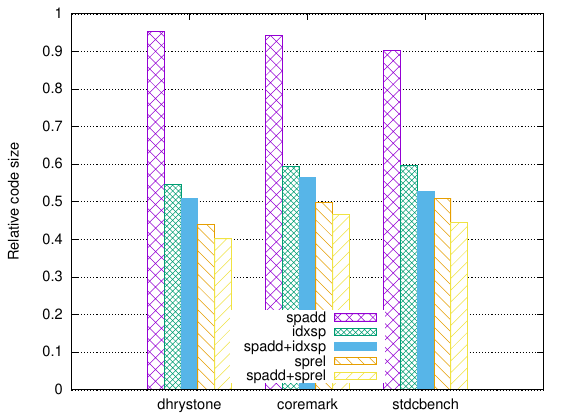}}
}
\caption{\label{results}Experimental results: Code size impact of instruction set modifications}
\end{figure*}

Figure~\ref{results} shows the results. All code sizes are relative to the compilation for the current Padauk architecture. Code sizes are shown for adding just \texttt{spadd \#k} (spadd), for adding stackpointer-relative addressing to \texttt{mov} only (idxsp), for adding both (spadd+idxsp), for adding stackpointer-relative addressing in all instructions (sprel) and for adding both this and \texttt{spadd \#k} (spadd+sprel).

The code size impact is much bigger when all functions are compiled as reentrant (code size reductions up to 60\%) than otherwise (code size reduction below 13\% in all cases). Also, for reentrant functions full stackpointer-relative addressing (sprel) provides a much bigger advantage over stackpointer-relative addressing for \texttt{mov} only (idxsp) than for other functions. Code size reduction from adding stackpointer-relative addressing increases from pdk13 over pdk14 to pdk15: For all architectures, the new  stackpointer-relative addressing covers a quarter for the address space, so the larger the address space, the more accesses to local variables can benefit from it. Since the program memory takes up about as much die space as the rest of the digital logic combined (Figure~\ref{PMC234-die}), what little area would be needed to implement the new functionality (essentially from a single 4-bit adder for pdk13 to a single 7-bit adder for pdk16) is more than made up for by savings in code memory requirements.

Still, even for spadd+sprel when compiling all functions as reentrant code size was still slightly higher than for the current Padauk architecture when compiling functions as non-reentrant. However, this is likely to change once bytewise register allocation is implemented in the compiler, which SDCC already supports when targeting the STM8~\cite{Bytewise}. Bytewise register allocation would allow keeping one or two bytes of a variable in \texttt{a} or \texttt{p} while spilling other bytes of the same variable.

Since full reentrancy is particularly useful for the multicore devices, we recommend full stackpointer-relative addressing and addition of \texttt{spadd} for those. For other devices, a stackpointer-relative \texttt{mov} variant might be sufficient.

For nested critical sections, the global interrupt enable bit should be exposed in an I/O register, so it can be accessed by the \texttt{swapc} atomic bit-swap instruction. This would also free the \texttt{engint} (enable interrupts) and \texttt{disgint} (disable interrupts) opcodes, as existing bit reset \texttt{set0} and bit set \texttt{set1} could be used instead.

To implement standard-conforming \texttt{atomic\_flag} efficiently, a swap instruction with indirect addressing mode is essential, which could be added in the form of \texttt{idxxch m, a} to all devices (sacrificing other instructions on pdk13 and pdk14).

Having additional lock-free atomics would be very useful, in particular for the multicore devices. To implement efficient atomic 8-bit types, an atomic compare-and-exchange with indirect addressing is essential. An instruction to get the number of the core, on which code is currently executed would also be very useful. Further efficiency gains for atomics could come via a compare-and-exchange with direct addressing, as well as atomic addition, and, or, exclusive or with indirect addressing.
While an atomic 16-bit-wide compare-and-exchange would also be very useful (as it would allow for atomic pointers to the generic address space), it would probably require bigger changes to the architecture. The \texttt{pushw m} instruction already supported by a few of the current pdk16 devices also helps with more efficient passing of arguments to reentrant functions. The \texttt{igoto} and \texttt{icall} instruction present in some older pdk16 devices make the use of function pointers more efficient.
Converting a binary number to a decimal string representation (as in the \texttt{printf} function family in the standard library) is commonly implemented via repeated division by 10. The Padauk µC do not have hardware support for division, and adding it would be expensive. However, this conversion can also be done efficiently using Binary Coded Decimal (BCD) arithmetic. Support for BCD could be added via a \texttt{da a} instruction to adjust the value in the accumulator after an addition or subtraction based on the existing carry and half carry flags.

\section{Conclusion}

We have implemented C for the tiny Padauk microcontrollers, and looked into possible improvements to make the hardware a better target for C. We present some small changes to the instruction set that would make the microcontrollers a much better target for C, which is also reflected in the impact on code size.

We thank the Padauk employees that helped us with questions on details of the hardware, those that reverse engineered and documented the opcodes, and the Easy PDK programmer developer.

\bibliographystyle{unsrt}
\bibliography{PdkC}

\end{document}